\documentclass[conference]{IEEEtran}
\usepackage[acronym,shortcuts,automake,nogroupskip,nopostdot]{glossaries}

\makeglossaries
\usepackage[table,dvipsnames]{xcolor}
\definecolor{domainblue}{RGB}{221,226,247}
\usepackage{graphicx}
\usepackage[utf8]{inputenc}
\usepackage{tabularx}
\usepackage{multirow}
\usepackage{booktabs}
\usepackage{amsmath}
\usepackage[most]{tcolorbox}
\usepackage{makecell}

\title{Threat Modelling using Domain-Adapted Language Models: Empirical Evaluation and Insights}

\author{
\IEEEauthorblockN{Saba Pourhanifeh\IEEEauthorrefmark{1},
AbdulAziz AbdulGhaffar\IEEEauthorrefmark{1}, 
Ashraf Matrawy\IEEEauthorrefmark{2}}
\IEEEauthorblockA{\IEEEauthorrefmark{1}Department of Systems and Computer Engineering}
\IEEEauthorblockA{\IEEEauthorrefmark{2}School of Information Technology}
Carleton University\\
\\
\IEEEauthorblockA{Emails: \{sabapourhanifeh, abdulazizabdulghaff\}@cmail.carleton.ca,}
\IEEEauthorblockA{AshrafMatrawy@cunet.carleton.ca}
{\bf Authors’ draft for soliciting feedback -- } \today }

\begin{document}
\newacronym{1g}{1G}{First Generation}
\newacronym{2g}{2G}{Second Generation}
\newacronym{3g}{3G}{Third Generation}
\newacronym{4g}{4G}{Fourth Generation}
\newacronym{5g}{5G}{Fifth Generation}
\newacronym{5g-advanced}{5G-Advanced}{5G-Advanced}
\newacronym{6g}{6G}{Sixth Generation}

\newacronym{vnf}{VNF}{Virtualized Network Function}

\newacronym{mmtc}{mMTC}{massive Machine Type Communications}
\newacronym{embb}{eMBB}{enhanced Mobile Broadband}
\newacronym{urllc}{URLLC}{Ultra-Reliable Low Latency Communications}
\newacronym{sdn}{SDN}{Software Defined Networking}
\newacronym{nfv}{NFV}{Network Function Virtualization}
\newacronym{nf}{NF}{Network Function}
\newacronym{dn}{DN}{Data Network}
\newacronym{arpf}{ARPF}{Authentication credential Repository and Processing Function}
\newacronym{soc}{SOC}{Security Operations Center}

\newacronym{pfcp}{PFCP}{Packet Forwarding Control Protocol}
\newacronym{udp}{UDP}{User Datagram Protocol}

\newacronym{supi}{SUPI}{Subscription Permanent Identifier}
\newacronym{capex}{CAPEX}{Capital Expenditure}
\newacronym{opex}{OPEX}{Operational Expenditure}
\newacronym{dos}{DoS}{Denial-of-Service}
\newacronym{ddos}{DDoS}{Distributed Denial-of-Service}
\newacronym{minlp}{MINLP}{Mixed-Integer Nonlinear Programming}
\newacronym{ilp}{ILP}{Integer Linear Programming}
\newacronym{upf}{UPF}{User Plane Function}
\newacronym{mec}{MEC}{Multi-access Edge Computing}
\newacronym{qos}{QoS}{Quality of Service}
\newacronym{sfc}{SFC}{Service Function Chain}
\newacronym{pdu}{PDU}{Protocol Data Unit}
\newacronym{ns}{NS}{Network Slicing}
\newacronym{sla}{SLA}{Service Level Agreement}
\newacronym{dps}{DPS}{Data Plane Services}
\newacronym{cps}{CPS}{Control Plane Services}
\newacronym{vm}{VM}{Virtual Machine}
\newacronym{mip}{MIP}{Mixed Integer Programming}

\newacronym{sba}{SBA}{Service-Based Architecture}

\newacronym{amf}{AMF}{Access and Mobility Function}
\newacronym{smf}{SMF}{Session Management Function}
\newacronym{nrf}{NRF}{Network Repository Function}
\newacronym{scip}{SCIP}{Solving Constraint Integer Programs}
\newacronym{ue}{UE}{User Equipment}

\newacronym{hplmn}{H-PLMN}{Home \gls{plmn}}
\newacronym{plmn}{PLMN}{Public Land Mobile Network}
\newacronym{eap-aka}{EAP-AKA'}{Extensible Authentication Protocol-Authentication and Key Agreement}
\newacronym{aka}{AKA}{Authentication and Key Agreement}
\newacronym{gnb}{gNB}{gNodeB}
\newacronym{3gpp}{3GPP}{3rd Generation Partnership Project}
\newacronym{ran}{RAN}{Radio Access Network}
\newacronym{udm}{UDM}{Unified Data Management}
\newacronym{ausf}{AUSF}{Authentication Server Function}

\newacronym{kpi}{KPI}{Key Performance Indicator}

%%%%%%%%%%%%%%%%%%%%%%%%%
\newacronym{llm}{LLM}{Large Language Model}
\newacronym{slm}{SLM}{Small Language Model}
\newacronym{genai}{GenAI}{Generative Artificial Intelligence}
\newacronym{ai}{AI}{Artificial Intelligence}
\newacronym{guti}{GUTI}{Globally Unique Temporary Identity}

\newacronym{imei}{IMEI}{International Mobile Equipment Identity}
\newacronym{s-nssai}{S-NSSAI}{Single Network Slice Selection Assistance Information} 
\newacronym{mitm}{MiTM}{Man-in-The-Middle}

\newacronym{stride}{STRIDE}{Spoofing, Tampering, Repudiation, Information Disclosure, Denial of Service, and Elevation of Privilege}

\newacronym{ltm}{LTM}{Large Telecom Model}

\newacronym{nssaa}{NSSAA}{Network Slice-Specific Authentication and Authorization}

\newacronym{zs}{ZS}{Zero-Shot}
\newacronym{fs}{FS}{Few-Shot}
\newacronym{mcq}{MCQ}{Multiple-Choice Question}
\newacronym{ior}{IOR}{Invalid Output Rate}
\newacronym{icl}{ICL}{In-Context Learning}

\newacronym{cp}{CP}{Control Plane}
\newacronym{up}{UP}{User Plane}

% \IEEEoverridecommandlockouts
% \makeatletter\def\@IEEEpubidpullup{6.5\baselineskip}\makeatother
% \IEEEpubid{\parbox{\columnwidth}{
% 		Network and Distributed System Security (NDSS) Symposium 2027\\
% 		22 - 26 March 2027, Seoul, South Korea\\
% 		ISBN 979-8-9919276-8-0\\  
% 		https://dx.doi.org/10.14722/ndss.2027.[23$|$24]xxxx\\
% 		www.ndss-symposium.org
% }
% \hspace{\columnsep}\makebox[\columnwidth]{}}

\maketitle

\begin{abstract}
\glspl{llm} are increasingly explored for cybersecurity applications such as vulnerability detection. In the domain of threat modelling, prior work has primarily evaluated a number of general-purpose \glspl{llm} under limited prompting settings. In this study, we extend the research area of {\it structured threat modelling} by systematically evaluating domain-adapted language models of different sizes to their general counterparts. We use both \glspl{llm} and \glspl{slm} that were domain adapted to telecommunications and cybersecuirty. For the structured threat modelling, we selected the widely used STRIDE approach and the application area is 5G security.

We present a comprehensive empirical evaluation using 52 different configurations (on 8 different language models) to analyze the impact of \textcircled{1} domain adaptation, \textcircled{2} model scale, \textcircled{3} decoding strategies (greedy vs. stochastic sampling), and \textcircled{4} prompting technique on STRIDE threat classification. Our results show that domain-adapted models do not consistently outperform their general-purpose counterparts, and decoding strategies significantly affect model behavior and output validity. They also show that while larger models generally achieve higher performance, these gains are neither consistent nor sufficient for reliable threat modelling. These findings highlight fundamental limitations of current \glspl{llm} for structured threat modelling tasks and suggest that improvements require more than additional training data or model scaling, motivating the need for incorporating more task-specific reasoning and stronger grounding in security concepts.

We present insights on invalid outputs encountered and present suggestions for prompting tailored specifically for STRIDE threat modelling.

\end{abstract}


%\tableofcontents

\section{Introduction}

\acrfullpl{llm} are increasingly explored for and integrated into a wide range of domains, including telecommunications and cybersecurity~\cite{zhou2024telecomsurvey,zhang2025llms,xu2024large,soman2023observations}. Their strong capabilities in natural language understanding, reasoning, and generalization have motivated a growing body of research investigating their suitability for diverse tasks. In the context of security, \glspl{llm} have the potential to assist practitioners in complex processes such as threat analysis, vulnerability assessment, and system-level reasoning~\cite{houssel2024towards}. 
Among these security tasks, structured threat modelling frameworks play a critical role, with \gls{stride}~\cite{Microsoftstride} serving as a widely adopted methodology for systematically identifying and categorizing security risks.
Despite this growing interest, there remains a limited understanding of how well \glspl{llm} perform in structured threat modelling tasks. 

Recent studies have examined the use of Language Models for various security tasks. One example of such work has evaluated general-purpose and code-specialized \glspl{llm} for software vulnerability detection in Java and C/C++ programs~\cite{lin2025large}. Lin \textit{et al.}~\cite{lin2025large} observe that performance is highly affected by model specialization, prompting strategy, and task formulation. In the context of threat modelling, AbdulGhaffar \textit{et al.}~\cite{abdulghaffar2025llmssuitabilitynetworksecurity} investigate the use of general-purpose proprietary \glspl{llm} for STRIDE threat modelling in 5G networks and similarly point out model sensitivity to different prompting strategies.

{\bf Domain-adapted \glspl{llm}:} Recent work has introduced domain-adapted \glspl{llm} trained on telecommunications and cybersecurity data, including models such as \texttt{Tele-LLMs}~\cite{maatouk2025telellmsseriesspecializedlarge}, \texttt{TelecomGPT}~\cite{zou2025telecomgpt}, and \texttt{Foundation-Sec}~\cite{kassianik2025llama31foundationaisecurityllmbase8btechnicalreport}. 
In this paper, we commonly use the term {\it Domain-adapted \glspl{llm}} to refer to models that are adapted to a specific domain through different training strategies, such as: full pretraining, domain-adaptive continual pre-training, fine-tuning, and instruction-tuning.
These models are typically obtained through continual pretraining on structured domain-specific corpora, such as standards documents, threat intelligence sources, and technical specifications, followed by instruction tuning to align model outputs with task requirements. In particular, models such as \texttt{Tele-LLMs}~\cite{maatouk2025telellmsseriesspecializedlarge} and \texttt{Foundation-Sec}~\cite{kassianik2025llama31foundationaisecurityllmbase8btechnicalreport} incorporate knowledge from telecommunications standards (e.g., \gls{3gpp}) and cybersecurity resources, making them promising candidates for various security tasks. However, their effectiveness for \gls{stride}-based threat modelling has not been systematically evaluated (see Section~\ref{domain}).

{\bf Scope: }This raises a key question that we address in this paper: \textit{can domain-adapted models perform \gls{stride}-based threat modelling?} 
In this work, we evaluate domain-adapted \glspl{llm} alongside their general-purpose counterparts for \gls{stride}-based threat classification. In addition to standard prompting-based evaluation, we investigate model behavior under stochastic sampling settings in order to capture output variability and assess consistency across multiple generations~\cite{rudd2025practicalguideevaluatingllms}. Furthermore, we focus on open-source, relatively small-scale \glspl{llm}, often referred to as \glspl{slm}~\cite{van2025survey}, enabling us to examine the interplay between model scale and domain specialization.

% In terms of chosen threats, we focus on 5G threats and vulnerabilities across different 5G interfaces. These threats were previously identified by Mahyoub et al.~\cite{mahyoub2024security}. Therefore, we chose six distinct threats covering all STRIDE categories for evaluating the models.

In this study, we focus on 5G threats and vulnerabilities associated with different \gls{5g} interfaces. We drew these threats from a study by Mahyoub \textit{et al.}~\cite{mahyoub2024security}, who identify threats across critical \gls{5g} interfaces and classify them using \gls{stride}. From this work, we select six threats that collectively cover all six STRIDE categories, enabling us to better evaluate models knowledge of \gls{stride}-based threat modelling.

This paper aims to answer the following research questions:

\textbf{RQ1 - Domain Adaptation:} Do domain-adapted language models trained on cybersecurity and telecommunications data demonstrate stronger task-specific knowledge than their general-purpose counterparts?

\textbf{RQ2 - Model Properties:} How do different decoding strategies and model size affect performance in STRIDE threat classification? What accuracy or response-quality differences are observed between smaller and larger models?

\textbf{RQ3 - Prompting Technique:} How do prompt engineering strategies, chat templates and \gls{fs} prompting, affect the quality, consistency, and accuracy of \gls{llm}-generated STRIDE threat classifications?

{\bf Contributions: }The main contributions of this paper are:

\textcircled{1} presenting a systematic evaluation of general-purpose and domain-adapted instruction-tuned \glspl{llm} for \gls{stride}-based threat classification. This analysis examines whether models adapted to cybersecurity or telecommunications domains demonstrate stronger task-specific performance than their general-purpose counterparts (Section~\ref{c1}). 

\textcircled{2} investigating the effect of model scale and decoding strategy on \gls{stride} classification performance. Specifically, we examine whether larger models consistently achieve better results and whether different decoding settings, including deterministic and stochastic sampling configurations, lead to consistent or improved classification outcomes (Section~\ref{c2}).

\textcircled{3} investigating the impact of prompting and formatting strategies on model performance and output quality. In particular, we compare \gls{zs} and \acrfull{fs} prompting, as well as the effects of utilizing chat templates where available, to better assess how each of these choices influences model performance, instruction-following and output quality (Section~\ref{c3}).

\textcircled{4} Section~\ref{c4} reports additional insights that emerged during the evaluation process and were not originally formulated as standalone research questions. These insights focus on unreliable output generation, including hallucinated content, invalid or out-of-scope labels, contradictory label and explanation pairs, empty or code-like responses, and cases where models failed to gracefully express uncertainty or abstain when a valid classification could not be determined.

\textcircled{5} Based on our observations on how prompting changes the results, we include prompting suggestions specific to structured threat modelling (Section~\ref{sec:prompting-guidlines}).

{\bf Paper organization:} The rest of this paper is structured as follows. Section~\ref{sec:related-work} discusses related work and identifies gaps. Section~\ref{sec:methodology} describes our methodology. Section~\ref{sec:evaluation} presents the experimental results and each model's performance under greedy decoding and stochastic sampling, followed by Section~\ref{sec:insights} that summarizes our main findings. We also include suggestions regarding prompting for threat modelling in Section~\ref{sec:prompting-guidlines}. We conclude our paper in Section~\ref{sec:conclusion}.

\section{Related Work}
\label{sec:related-work}

\subsection{Domain-adapted Models}
\label{domain}
Recent work has explored domain adaptation of \glspl{llm} for specialized technical domains, including telecommunications and cybersecurity. For instance, \texttt{Tele-LLMs}~\cite{maatouk2025telellmsseriesspecializedlarge} introduce the first open-source family of \glspl{llm} tailored specifically for the telecommunications domain. The authors construct \textit{Tele-Data}, a large-scale data curated by 1) \gls{llm}-based filtering of scientific arXiv publications, 2) \gls{3gpp} standards documents, 3) telecommunications-related Wikipedia articles, and 4) relevant web content collected through web-crawling. Using this dataset, the models are adapted through continual (domain-adaptive) pretraining~\cite{parmar2024reusedontretrainrecipe,yildiz2025investigatingcontinualpretraininglarge} from general-purpose pretrained \glspl{llm}. A subset of these models were further adopted by leveraging instruction tuning (Full Fine Tuning) using Alapaca and Open-instruct, which are referred to as \texttt{Tele-it} models.

For evaluation, the authors introduce \textit{Tele-Eval}, a benchmark consisting of approximately 750K telecom-specific open-ended question–answer pairs. Notably, this dataset is generated by prompting an \gls{llm}, \texttt{Mixtral-8x7B-Instruct}~\cite{mistral_mixtral_8x7b_instruct_modelcard}, with \textit{Tele-Data}~\cite{maatouk2025telellmsseriesspecializedlarge}. The authors deliberately avoid \gls{mcq} formats (e.g., TeleQnA~\cite{maatouk2025teleqna}), arguing that \glspl{llm} exhibit selection biases that can influence their ability as \gls{mcq} predictors~\cite{zheng2024largelanguagemodelsrobust}.
%Furthermore, they show that adapting a single unified model across multiple telecommunications sub-domains yields better overall performance than training narrowly specialized models, due to positive cross-domain knowledge transfer.

Another example is \texttt{TelecomGPT}~\cite{zou2025telecomgpt}, built through three standard adaptation stages, domain-specific continual pretraining, instruction and alignment tuning. The authors curated a telecom-specific dataset, named \textit{OpenTelecom}, consisting of 1) 3GPP standards, 2) IEEE standards, 3) research papers from arXiv, 4) related Wikipedia entries, etc. (However, they did not proceed to do this pre-training on \texttt{Llama-3-8B}~\cite{llama3modelcard} since it has already been trained on 15TB of tokens.) Their instruction tuning was performed using QLoRA (Quantized Low-Rank Adaptation) technique with \textit{TelecomInstruct} dataset, which trains the model further to perform critical tasks such as 1) \gls{mcq} Answering, 2) Open-ended Question Answering, 3) Technical document classification, 4) Math Modelling, 5) Code Generation and 6) Code Infilling. 

Similar efforts have investigated adapting general-purpose \glspl{llm} to the cybersecurity domain. Cisco introduced \texttt{Foundation-Sec-8B} \cite{kassianik2025llama31foundationaisecurityllmbase8btechnicalreport}, a cybersecurity-focused variant of \texttt{Llama-3.1} trained via continual pretraining on curated security corpora, followed by the release of \texttt{Foundation-Sec-8B-Instruct}~\cite{weerawardhena2025llama31foundationaisecurityllm8binstructtechnicalreport} to restore instruction-following capabilities comparable to \texttt{Llama-3.1-8B-Instruct}~\cite{llama318modelcard} while retaining domain-specific knowledge. The primary objective of this effort was to inject foundational cybersecurity knowledge into a general \gls{llm}, enabling downstream adoption across a range of security-related tasks such as \gls{soc} acceleration, threat intelligence analysis, and engineering support.

It is important to note that these domain-adapted models were not explicitly designed for structured threat modelling frameworks such as \gls{stride}. Nevertheless, given that \texttt{Foundation-Sec-8B-Instruct}~\cite{kassianik2025llama31foundationaisecurityllmbase8btechnicalreport} is trained on core cybersecurity standards, threat intelligence sources, and vulnerability documentation, it provides a compelling candidate for exploratory evaluation against structured threat modelling tasks. This also goes for \texttt{Tele-LLM}~\cite{maatouk2025telellmsseriesspecializedlarge} models since they were trained on 3GPP and related telecom documentations. Therefore, in this work, we include these models alongside their general-purpose counterparts to assess how domain-adaptive pretraining influences performance on \gls{stride}-based threat classification, despite the task not being an explicit training objective.

\subsection{Security evaluation using \glspl{llm}}
\label{llm-sec}
Lin \textit{et al.}~\cite{lin2025large} contribute valuable insights where they compare and evaluate various \glspl{llm} for detecting vulnerabilities in code written with Java and C/C++ programming languages. The authors study the impact of multiple factors on the performance of \glspl{llm} and highlight the key insights. Of the key takeaways, the authors noted that the performance of domain-adapted (`specialized') models is not consistently better than the general models. Furthermore, the \gls{fs} experiments in their work reveal surprisingly unexpected results by degrading performance compared to \gls{zs} experiments. Although the authors acknowledge these unexpected results and point out that these results might be due to the prompt engineering approach they used in their experiments.

AbdulGhaffar \textit{et al.}~\cite{abdulghaffar2025llmssuitabilitynetworksecurity} investigate the use of \glspl{llm} for \gls{stride}-based threat classification in \gls{5g} networks, evaluating several general-purpose proprietary models under \gls{zs} and \gls{fs} prompting settings~\cite{abdulghaffar2025llmssuitabilitynetworksecurity}. Their study highlights several important limitations, including (i) incorrect threat perspective, where models misinterpret the viewpoint of the threat, (ii) difficulty in identifying secondary threats, and (iii) sensitivity to prompting strategies, with \gls{fs} prompting generally improving performance. Based on these findings, the authors emphasize the need for adapting or fine-tuning \glspl{llm} for network security use cases.

\subsection{Gap in Related Work} 
\textbf{Vulnerability detection:} In this paper, we focus on a different scope and take a distinct approach compared to the work by Lin \textit{et al.}~\cite{lin2025large} where they focus on utilizing \glspl{llm} to identify vulnerabilities in programming language codes. We use \glspl{llm} to classify \gls{5g} threats and vulnerabilities based on \gls{stride} threat model. Additionally, their approach~\cite{lin2025large} differs from our work as they utilize \glspl{llm} that are specifically tailored for code-related tasks (e.g., \texttt{CodeGemma}, \texttt{CodeLLaMA}) along with general-purpose \glspl{llm} in their evaluation. In this work, we use cybersecurity and telecommunications domain-adapted \glspl{llm} along with their general-purpose counterparts in our evaluation.

\textbf{\gls{llm}-based \gls{stride} classification:} The work of AbdulGhaffar \textit{et al.} \cite{abdulghaffar2025llmssuitabilitynetworksecurity} on \gls{llm}-based \gls{stride} classification primarily focuses on studying the suitability of general-purpose proprietary \glspl{llm} for \gls{stride} modelling. The major limitation of their work lies in not considering the telecom or security domain-specific \glspl{llm} for evaluation. In this work, our analysis of the domain-specific \glspl{llm} reveals novel and valuable insights on the ability of these models in \gls{stride} threat modelling tasks.

Overall, existing work has either focused on general-purpose proprietary \glspl{llm} for STRIDE-based threat modelling~\cite{abdulghaffar2025llmssuitabilitynetworksecurity} or on evaluating domain-adapted models on different telecommunications and cybersecurity benchmarks~\cite{zou2025telecomgpt, maatouk2025telellmsseriesspecializedlarge,kassianik2025llama31foundationaisecurityllmbase8btechnicalreport}. Therefore, the effect of domain adaptation, model scale, prompt formatting, and decoding strategy on STRIDE-based \gls{5g} threat classification remains underexplored.

\section{Methodology}
\label{sec:methodology}
This section describes the methodology and experimental setting used to evaluate language models for \gls{stride}-based threat classification in \gls{5g} network security. This task is formulated as a multi-label classification problem, where each threat may be associated with one or more \gls{stride} categories. The evaluation is designed to compare general-purpose and domain-adapted instruction-tuned models in different scales under different prompting and decoding conditions. We first describe the task formulation and the selected threats, then present the prompt structure, evaluated models, generation settings, output validation procedure, evaluation metrics and finally the execution environment. 

\subsection{Task Setup}

Consistent with the multi-label nature of STRIDE-based threat modelling, each threat in this study was allowed to correspond to one or more STRIDE categories. This is important because a single threat may involve multiple security properties, such as information disclosure together with spoofing or tampering under certain conditions. Therefore, evaluation was performed at the level of individual STRIDE categories rather than as a mutually exclusive multi-class classification problem.

This setup also avoids forcing the model into a single-label decision when multiple categories may be applicable. Prior work has noted that constrained answer formats, such as multiple-choice or single-label settings, can introduce selection bias in \gls{llm} outputs~\cite{zheng2024largelanguagemodelsrobust, maatouk2025telellmsseriesspecializedlarge,ma-etal-2025-large-language}. This concern is particularly relevant for threat modelling, where secondary but still valid STRIDE categories may be suppressed if the model is required to select only one dominant label.

\begin{table*}[h]
    \centering
    \caption{Representative 5G threat descriptions provided as input to the models.}
    \label{tab:threat-list}
    \begin{tabular}{|p{6.2cm}|p{11cm}|}\hline
         \cellcolor{domainblue} \textbf{5G Threats} & \cellcolor{domainblue} \textbf{Threat Description} \\\hline
          AMF impersonation on N1 interface & An attacker may run a rogue \gls{amf} and serve the \glspl{ue}, resulting in the  disclosure of sensitive user information~\cite{mahyoub2024security,3GPP.33.501,3GPP.33.926} \\\hline
          5G-GUTI and IMEI correlation on N1 interface & The location of the \gls{ue} can be revealed if the attacker is able to correlate the \gls{imei} and \gls{guti} of the victim \gls{ue}~\cite{mahyoub2024security, nist_5g} \\\hline
          Bidding down on Xn-handover & This threat occurs during Xn-based handover when the \gls{amf} does not verify the security capabilities of the \gls{ue} exchanged from the source \gls{gnb} to the target \gls{gnb}. Even when the \gls{ue} supports secure network protocols, the source \gls{gnb} can maliciously force the \gls{5g} network to use less secure protocols (bid down), leaving the \gls{ue} vulnerable to attacks~\cite{mahyoub2024security, 3GPP.33.926} \\\hline
          Eavesdropping on F1 interface & The \gls{cp} and \gls{up} traffic on the F1 interface can be eavesdropped, which may result in information disclosure. This can further result in spoofing and tampering if the eavesdropped traffic is replayed~\cite{mahyoub2024security,3GPP.33.926, 3GPP.33.824} \\\hline
          False \gls{s-nssai} on N1 interface & A malicious \gls{ue} may send a malformed \gls{s-nssai} during the network slice authentication procedure, resulting in the escalation of privileges by joining an unauthorized slice~\cite{3GPP.33.501, 3GPP.33.926}  \\\hline
          Man-in-The-Middle (MiTM) attack on N3 interface & This attack can occur between the \gls{gnb} and the \gls{upf} if proper security measures are not implemented~\cite{mahyoub2024security}\\\hline
    \end{tabular}
\end{table*}

\subsection{Selected \acrshort{5g} Threats}
We select six \gls{5g} threats on various \gls{5g} network interfaces and their \gls{stride} classifications from the study of Mahyoub \textit{et al.}~\cite{mahyoub2024security}. In their work,  Mahyoub \textit{et al.}~\cite{mahyoub2024security} extensively identify the threats and vulnerabilities on critical \gls{5g} network interfaces and classify them using \gls{stride} threat model. In our study, we opted for six threats, as illustrated in Table~\ref{tab:threat-list}, only to be able to manually evaluate all experiments and not rely on automated evaluation (e.g., by \glspl{llm}) and to have control over the evaluation and remove the impact of \gls{llm} inaccuracy on the evaluation.

The six selected threats represent all six categories of \gls{stride} threat model and diversity of \gls{5g} network interface representation. This approach allows us to analyze the performance of \glspl{llm} in \gls{stride} threat classification. A brief description of each selected threat is given in Table~\ref{tab:threat-list}.

\subsection{Prompt Structure}
All models were prompted using a unified chat-based format. A short system prompt defined the model’s role as a 5G network security expert, while the user prompt specified the \gls{stride} classification task and concluded with the instruction \emph{“Classify the following threat or vulnerability:”}, followed by the threat description.
This prompt structure was kept identical across all experiments to ensure fair comparison.
The prompt formulation, \gls{stride} category definitions, and threat descriptions were adopted from prior work~\cite{abdulghaffar2025llmssuitabilitynetworksecurity}.

\begin{figure*}[t]
    \centering
    \fbox{
    \begin{minipage}{0.95\textwidth}
    \footnotesize
    \setlength{\parskip}{2pt}
    
    \textcolor{RoyalBlue}{\textbf{System Prompt:}}
    You are a 5G network security expert. Your task is to classify a given 5G threat or vulnerability according to the STRIDE model:
    1. Spoofing
    2. Tampering
    3. Repudiation
    4. Information disclosure
    5. Denial of service
    6. Elevation of privilege
    Each threat or vulnerability may belong to one or more STRIDE categories.
    Your response should list only the applicable category or categories without any additional details or explanations.
    
    \vspace{0.3em}
    \textcolor{ForestGreen}{\noindent\textbf{User Prompt (Zero-shot):}}
    Classify the following threat/vulnerability:
    
    \vspace{0.3em}
    \textcolor{BrickRed}{\noindent\textbf{Threat:}} 
    \emph{[Threat description]}
    \end{minipage}%
    }
    \caption{Zero-shot STRIDE classification prompt used across all evaluated models.}
    \label{fig:zero-shot-prompt}
\end{figure*}

\begin{figure*}[t]
    \centering
    \fbox{
    \begin{minipage}{0.95\textwidth}
    \footnotesize
    \setlength{\parskip}{2pt}
    
    \textcolor{RoyalBlue}{\textbf{System Prompt:}}
    You are a 5G network security expert.
    Your task is to classify a given 5G threat or vulnerability according to the STRIDE model:
    1. Spoofing
    2. Tampering
    3. Repudiation
    4. Information disclosure
    5. Denial of service
    6. Elevation of privilege
    Each threat or vulnerability may belong to one or more STRIDE categories.
    Your response should list only the applicable category or categories without any additional details or explanations.

    \vspace{0.3em}
    \textcolor{ForestGreen}{\noindent\textbf{User Prompt (Few-Shot):}}
    Classify the following threat/vulnerability:
    
    The following are some examples of threat STRIDE classification.
    
            Here, \{X\} represents that the threat does not belong to this category, and \{O\} means the threat belongs to this category: 
            
            1. NAS protocol-based attack on N1 interface: S\{X\}, T\{X\}, R\{X\}, I\{O\}, D\{O\}, E\{X\}
            
            2. A bidding down of Security features on N1 interface: S\{X\}, T\{O\}, R\{X\}, I\{O\}, D\{O\}, E\{X\}
            
            3. Keystream reuse on Xn interface: S\{X\}, T\{X\}, R\{X\}, I\{O\}, D\{X\}, E\{X\}
            
            4. Flawed Validation of Client Credentials Assertion on SBI interface: S\{O\}, T\{X\}, R\{X\}, I\{O\}, D\{O\}, E\{O\}

    \vspace{0.3em}
    \textcolor{BrickRed}{\noindent\textbf{Threat:}} 
    \emph{[Threat description]}
    \end{minipage}%
    }
    \caption{Few-shot STRIDE classification prompt with in-context demonstrations used across all evaluated models.}
    \label{fig:few-shot-prompt}
\end{figure*}

For \gls{zs} prompting, models were evaluated without providing any task-specific examples, relying solely on task instructions and the input threat description. The exact prompts used in the \gls{zs} and \gls{fs} experiments are shown in Figures~\ref{fig:zero-shot-prompt} and~\ref{fig:few-shot-prompt}, respectively. We note that the prompt we use in \gls{fs} experiments does not include an example of a threat that belongs to the `repudiation' \gls{stride} category. This is due to relying on Mahyoub \textit{et al.}~\cite{mahyoub2024security} to obtain the \gls{5g} threats for \gls{llm} evaluation and the examples for \gls{fs} prompt. They~\cite{mahyoub2024security} only provide one threat that belongs to the `repudiation' category and we use this threat (\gls{mitm} attack on N3 interface) as one of the six threats to evaluate \glspl{llm}.

\subsection{Models Evaluated} %should cite all papers of these models here.

The selected models for evaluation are presented in table~\ref{tab:evaluated-models} with detailed discussion in Section~\ref{domain}.
It is worth noting that all evaluated models were instruction-tuned.
When chat templates were available, models were evaluated both with and without the corresponding template to assess their impact on prompt interpretation.
For models without available chat templates, the \texttt{Tele-it} models~\cite{maatouk2025telellmsseriesspecializedlarge}, prompts were provided in a plain text format due to architectural constraints.

\begin{table}[!t]
\centering
\caption{Summary of evaluated models. The domain-adapted group includes telecom-adapted \texttt{Tele-it}~\cite{maatouk2025telellmsseriesspecializedlarge} models and the cybersecurity-adapted \texttt{Foundation-Sec-8B-Instruct}~\cite{weerawardhena2025llama31foundationaisecurityllm8binstructtechnicalreport} model; detailed adaptation and pretraining information is discussed in the related work section.}
\label{tab:evaluated-models}
\renewcommand{\arraystretch}{1.25}
\setlength{\tabcolsep}{3pt}
\footnotesize

\begin{tabularx}{\columnwidth}{
>{\centering\arraybackslash}m{1.2cm}
>{\raggedright\arraybackslash}X
>{\centering\arraybackslash}m{0.7cm}
>{\centering\arraybackslash}m{1.2cm}
>{\centering\arraybackslash}m{1.4cm}
}
\toprule
\textbf{Category} 
& \textbf{Model Name} 
& \textbf{Size} 
& \textbf{Domain} 
& \makecell{\textbf{Chat Template?}} \\ 
\midrule

\multirow[c]{4}{1.5cm}{General-\\purpose}
& Meta-Llama-3-8B-Instruct~\cite{llama3modelcard} 
& 8B 
& General 
& Yes \\

& Llama-3.1-8B-Instruct~\cite{llama318modelcard}
& 8B 
& General 
& Yes \\

& Llama-3.2-3B-Instruct~\cite{llama323modelcard}
& 3B 
& General 
& Yes \\

& Llama-3.2-1B-Instruct~\cite{llama321modelcard} 
& 1B 
& General 
& Yes \\

\midrule

\multirow[c]{4}{1.5cm}{Domain-\\adapted}
& Llama-3-8B-Tele-it~\cite{maatouk2025telellmsseriesspecializedlarge} 
& 8B 
& Telecom 
& No \\

& Foundation-Sec-8B-Instruct~\cite{weerawardhena2025llama31foundationaisecurityllm8binstructtechnicalreport} 
& 8B 
& Cybersecurity 
& Yes \\

& Llama-3.2-3B-Tele-it~\cite{maatouk2025telellmsseriesspecializedlarge} 
& 3B 
& Telecom 
& No \\

& Llama-3.2-1B-Tele-it~\cite{maatouk2025telellmsseriesspecializedlarge}  
& 1B 
& Telecom 
& No \\

\bottomrule
\end{tabularx}
\end{table}

For clarity, in this work we use the term \texttt{Tele-LLMs} to refer to the model family introduced by Maatouk \textit{et al.}, while \texttt{Tele-it} refers to their instruction-tuned models evaluated in this study~\cite{maatouk2025telellmsseriesspecializedlarge}. Also, in Figures~\ref{fig:greedy-marked} and~\ref{fig:sampling-stride-classification}, as well as Tables~\ref{tab:greedy-metrics} and~\ref{tab:sampling-metrics}, each evaluated model is reported using its Hugging Face repository identifier in the \texttt{author/model} format.

\subsection{Experimental Setup}

To ensure consistency across all experimental settings, and given that the task required only label outputs without explanations, the maximum number of generated tokens was limited to \textbf{100} tokens. This limit provided sufficient space for multi-label outputs while preventing unnecessarily long responses, which was important since all model outputs were manually reviewed during evaluation.

In order to assess both deterministic classification performance and response consistency under controlled randomness~\cite{rudd2025practicalguideevaluatingllms}, each model configuration was evaluated using two decoding strategies: (i) \emph{Greedy decoding}, where sampling was disabled (\texttt{do\_sample=False}), and 
(ii) \emph{Stochastic sampling}, where sampling was enabled (\texttt{do\_sample=True}).

In greedy decoding, parameters such as \texttt{temperature} and \texttt{top\_p} were automatically ignored; however, in sampling-based experiments, the \texttt{temperature} and \texttt{top\_p} were set to \textbf{0.7} and \textbf{0.9}, respectively. 
Previous studies~\cite{renze-2024-effect,li2025exploring} explain how \texttt{temperature} affects models of varying sizes differently, with smaller models showing higher sensitivity to \texttt{temperature} increases, while larger models demonstrate better robustness across sampling settings. Since our study evaluates relatively small language models in a constrained multi-label classification task, higher temperatures were avoided. In addition, nucleus sampling with \texttt{top\_p} = 0.9 was used to limit token selection to a high-probability subset of the model's output distribution~\cite{Holtzman2020The}. Unlike open-ended or creative generation tasks, \gls{stride} threat classification does not require creativity; instead, it requires the model to select appropriate labels from a predefined label space. Excessive randomness may therefore increase the likelihood of invalid outputs, unsupported explanations, hallucinated labels, or inconsistent multi-label predictions.

Finally, as this study focuses on comparing model behavior under deterministic and sampling-based settings, we did not evaluate beam-search decoding. Since beam search maintains multiple candidate sequences, its impact on structured multi-label \gls{stride} classification may differ from greedy or sampling methods and we leave this investigation for future work. The generation parameters used in our experiments are summarized in Table~\ref{tab:generation-settings}.

\begin{table}[h]
    \centering
    \caption{Generation parameters for greedy decoding and stochastic sampling}
    \label{tab:generation-settings}
    \begin{tabular}{|c|c|c|}\hline
         \cellcolor{domainblue} Decoding Setting & \cellcolor{domainblue} Greedy Decoding & \cellcolor{domainblue} Stochastic Sampling \\\hline
         Number of runs &  1 & 10\\\hline
         \texttt{do\_sample} &  False & True\\\hline
         \texttt{max-new-tokens} &  100 & 100\\\hline
         \texttt{num\_beams} &  1 & 1\\\hline
         \texttt{temperature} &  Not applied & 0.7\\\hline
         \texttt{top\_p} &  Not applied & 0.9\\ \hline
    \end{tabular}
\end{table}

For stochastic decoding, each threat was evaluated over 10 independent runs using the same prompt and generation settings. Since stochastic sampling might produce different outputs across runs, we evaluated the outputs based on \gls{stride} category selection frequency. For each category, we computed the proportion of runs in which that category was selected by the model~\cite{zheng2024largelanguagemodelsrobust}. 
A category was considered chosen for a given threat if it appeared in at least \textbf{50\%} of the runs, corresponding to a total of 5 or more out of 10 runs. In a few cases where no category reached this threshold, category or categories with the highest selection frequency were chosen to ensure that each threat received at least one final \gls{stride} classification label. The aggregated labels were then used for metric computation. On the other hand, greedy decoding was executed once per threat due to its deterministic nature. 

\subsection{Output Validation and Evaluation Metrics}

For each evaluated model, prompting setting, chat-template condition, and decoding strategy, model outputs were recorded and organized according to their corresponding experimental configuration. The collected outputs were then manually reviewed to identify invalid, empty, or internally inconsistent responses.

After this review, \gls{stride} labels were manually assigned for each threat under each experimental configuration. All validation and annotation steps were performed by the authors without automated tools or AI-assisted labelling, ensuring full human oversight of the evaluation process.

Based on the finalized annotations, true positives (TP), false positives (FP), true negatives (TN), and false negatives (FN) were computed for each \gls{stride} category. These values were then used to calculate accuracy, precision, recall, and F1-score. We mainly focus on the F1-score since the task requires minimizing both false positives and false negatives. 

In addition to standard classification metrics, we computed the \gls{ior} to capture cases where model outputs failed to conform to the \gls{stride} classification task. \gls{ior} is computed as the number of invalid outputs divided by the total number of outputs for a given experimental configuration, as shown in Eq.~\ref{eq:ior}. Some examples of invalid outputs are: empty responses, outputs without usable \gls{stride} labels, hallucinated or non-STRIDE categories, code generation without label generation, conflicting classifications, and responses where the explanation contradicted the final label.

\begin{equation}
    \text{Invalid Output Rate (IOR)} = \frac{\text{Invalid Outputs}}{\text{Total Outputs}}
    \label{eq:ior}
\end{equation}

To account for both classification performance and output reliability, we used penalized F1-score as the primary metric:

\begin{equation}
    \text{Penalized F1-Score} = \text{F1-Score} \times \left(1 - \text{IOR}\right)
    \label{eq:penalF1}
\end{equation}

This metric penalizes models that achieve reasonable label-level performance but frequently fail to follow the required output format.

\subsection{Execution Environment}
All experiments were conducted on a single NVIDIA A100 80GB PCIe GPU. Models were executed in evaluation mode using \texttt{model.eval()} and \texttt{bfloat16} precision, with no gradient computation and no updates to model weights. Inference was performed in a dedicated Python 3.10.11 virtual environment with PyTorch 2.5.1 compiled with CUDA 11.8 support. Models and tokenizers were loaded using the Hugging Face Transformers library~\cite{huggingface}.

\section{Evaluation}
\label{sec:evaluation}

This section presents the results of our evaluation of the selected language models on the \gls{stride}-based threat classification task. We organize the results according to the two decoding settings used in the experiments: 1) greedy decoding and 2) stochastic sampling. We use penalized F1 (Eq.~\ref{eq:penalF1}) as the main metric because \gls{stride}-based threat classification requires not only accurate category selection, but also reliable adherence to the expected output format.

For example, a response may describe a threat as Spoofing but output Tampering as the final category, or it may incorrectly redefine one \gls{stride} category as another. Such outputs are treated as invalid because they cannot be reliably used in our threat-modelling workflow. Examples of such invalid outputs are available in the Appendix (Section~\ref{sec:appendix}). 

We first present the greedy-decoding results, where each model produces a single deterministic output for each threat description. We then present the sampling-based results, where each model is evaluated across repeated runs to capture output variability and category-selection consistency. Finally, we analyze invalid outputs across both decoding settings to better understand model reliability, instruction adherence, and failure patterns beyond standard classification performance.

\subsection{Greedy Decoding}
In the greedy setting, each model configuration produces one deterministic output for each threat description. Therefore, the results reflect each models highest probability prediction and does not consider any variability. By evaluating the greedy results, we asses how model size, model family (base architecture), prompting strategy (\gls{zs} and \gls{fs}), using chat-template, and domain adaptation affect \gls{stride} classification.

\begin{figure*}[!t]
    \centering
    \includegraphics[width=0.95\textwidth]{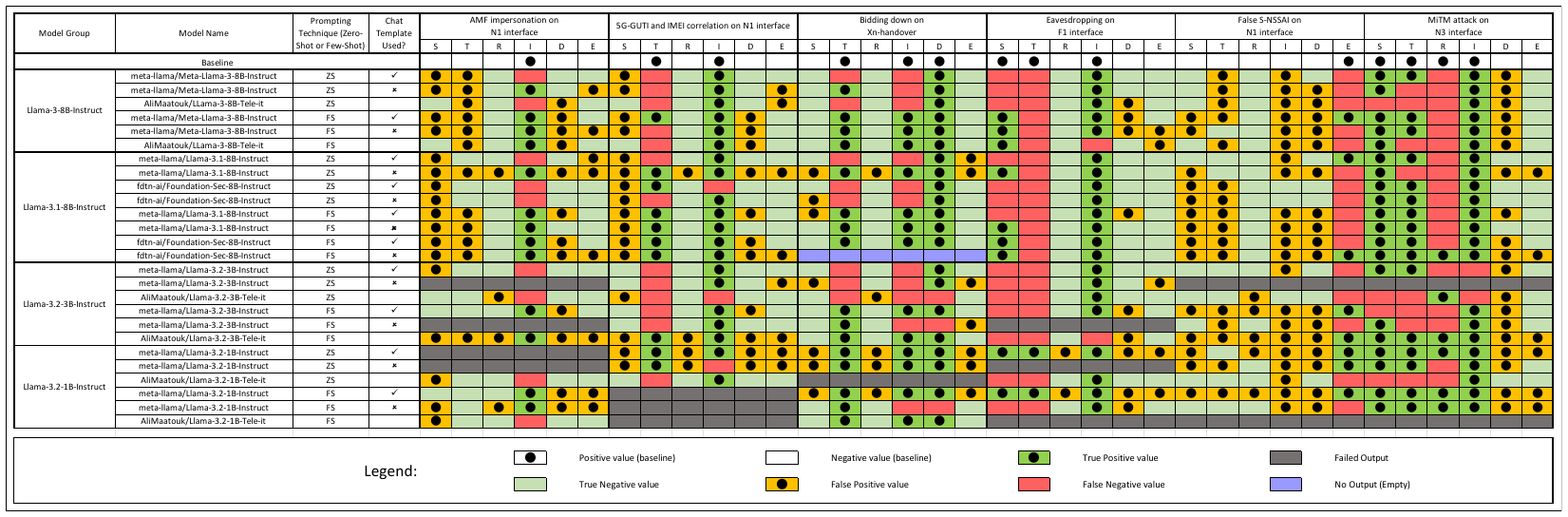}
    \caption{STRIDE Classification of 5G Threats under Greedy Decoding.}
    \label{fig:greedy-marked}
\end{figure*}

Figure~\ref{fig:greedy-marked} presents every model's \gls{stride} label prediction for the selected \gls{5g} threat descriptions, while Table~\ref{tab:greedy-metrics} reports the corresponding metrics. Since this task requires both correct classification of \gls{stride} labels and also adherence to the instructions, the main chosen metric is the penalized F1-score. 

\begin{table*}[!t]
    \centering
    \caption{Metrics Calculated based on Greedy Decoding Results. \\ \scriptsize{Note: Different cell colors indicate relative metric quality. \textcolor{Green}{\textbf{Green}} represents more desirable values, while \textcolor{red}{\textbf{Red}} represents less desirable values and white shows neutral results.}}
    \label{tab:greedy-metrics}
    \includegraphics[width=0.95\textwidth]{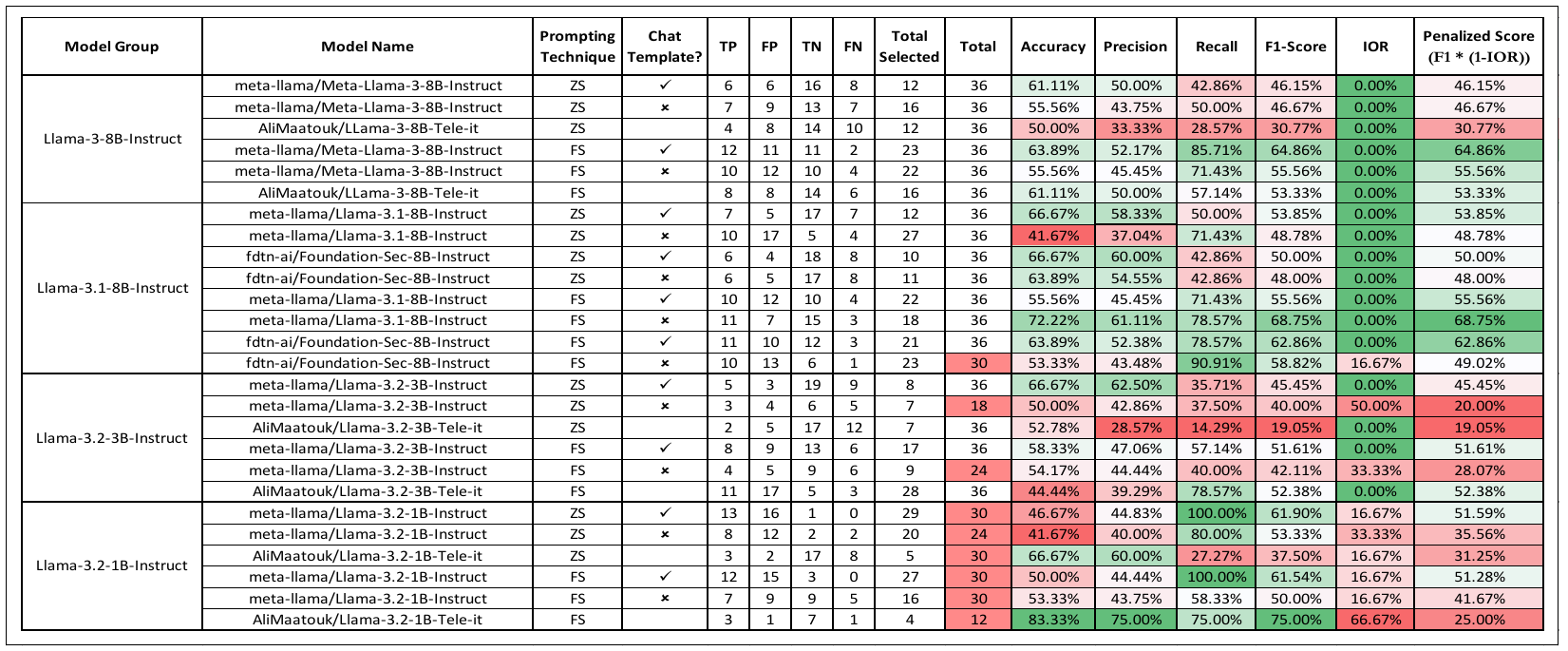}
\end{table*}

Overall, the greedy results show substantial variation across model configurations. The highest penalized F1-score was achieved by \texttt{Llama-3.1-8B-Instruct}~\cite{llama318modelcard} under \gls{fs} prompting without the chat template, reaching 68.75\%. Other high performing models include \texttt{Meta-Llama-3-8B-Instruct}~\cite{llama3modelcard} with \gls{fs} prompting and chat-template usage, achieving 64.86\%, and \texttt{Foundation-Sec-8B-Instruct}~\cite{weerawardhena2025llama31foundationaisecurityllm8binstructtechnicalreport} with \gls{fs} prompting and chat-template usage, which achieved 62.86\%. This suggests that the highest penalized-F1 scores in greedy decoding setting were generally achieved by 8B-parameter models under \gls{fs} prompting, therefore, larger model scale and in-context examples are important contributors to \gls{stride} classification performance.

\begin{figure}[!t]
    \centering    
    \includegraphics[width=0.95\linewidth, height=5.2cm]{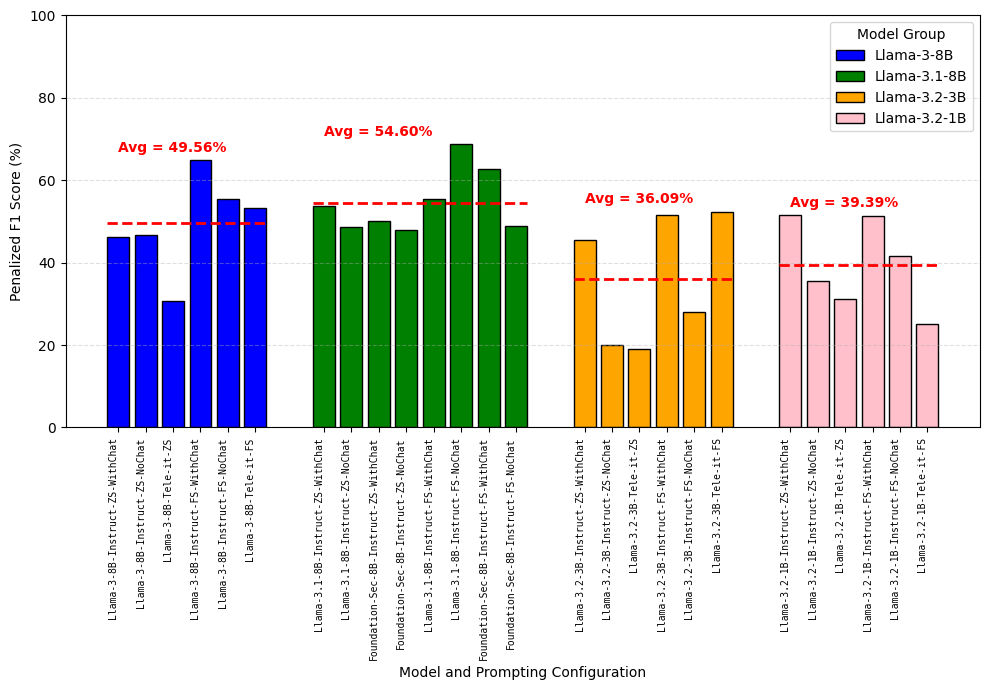}
    \caption{Penalized F1-Score under Greedy Decoding}
    \label{fig:f1-greedy}
\end{figure}

We also analyze the results based on each model group. Figure~\ref{fig:f1-greedy} shows that the larger models generally outperformed the smaller ones. The highest model group average penalized-F1 score was observed for the \texttt{Llama-3.1-8B}~\cite{llama318modelcard} group at 54.60\%, followed by \texttt{Llama-3-8B}~\cite{llama3modelcard} at 49.56\%. The smaller \texttt{Llama-3.2-1B}~\cite{llama321modelcard} and \texttt{Llama-3.2-3B}~\cite{llama321modelcard} groups achieved lower average penalized F1 scores of 39.39\% and 36.09\%, respectively. This shows that larger models generally perform better under deterministic decoding, however, the relationship is not strictly monotonic.

The results also show that domain adaptation alone does not guarantee better performance. For example, \texttt{Foundation-Sec-8B-Instruct}~\cite{weerawardhena2025llama31foundationaisecurityllm8binstructtechnicalreport} achieved comparable performance in some \gls{fs} settings, but it did not consistently outperform its general-purpose \texttt{Llama-3.1-8B-Instruct}~\cite{llama318modelcard} counterpart. Similarly, the \texttt{Tele-it}~\cite{maatouk2025telellmsseriesspecializedlarge} models showed mixed results across model sizes and prompting settings. 

\newtcolorbox{rqbox0}{
    colback=white,
    colframe=black,
    boxrule=0.5pt,
    arc=1pt,
    left=4pt,
    right=4pt,
    top=4pt,
    bottom=4pt,
    fonttitle=\bfseries,
    before skip=6pt,
    after skip=6pt
    }

\begin{rqbox0}
These results may indicate that domain-adaptive training reaches a point of diminishing returns when the underlying general-purpose model has already acquired substantial cybersecurity and telecommunications knowledge during large-scale pretraining. In smaller models, the limited model capacity may further restrict the extent to which additional domain-specific training can be converted into better task-level performance. Therefore, the results suggest that domain adaptation is not sufficient on its own.
\end{rqbox0}

% width=0.95\columnwidth
% width=0.95\linewidth, height=6cm

\begin{figure}[!t]
    \centering    
    \includegraphics[width=0.95\columnwidth]{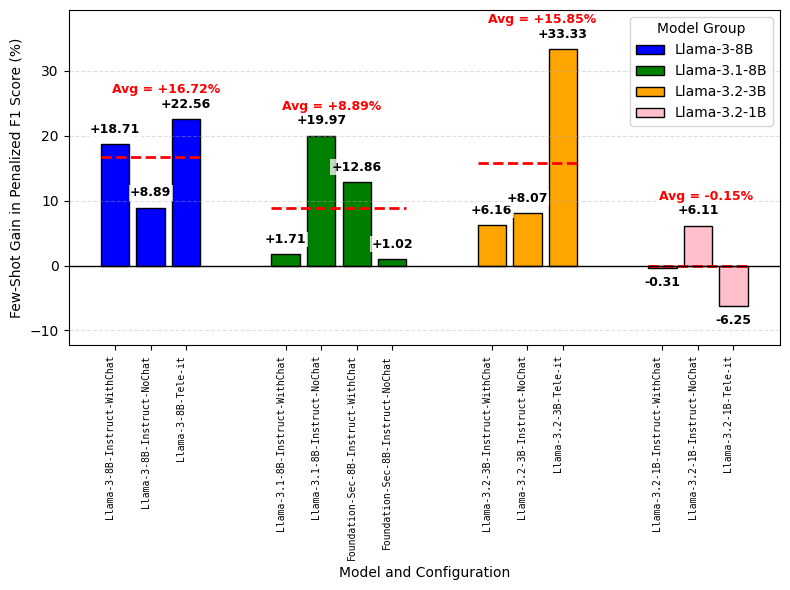}
    \caption{Few-shot Performance Gain under Greedy Decoding}
    \label{fig:fs-gain-greedy}
\end{figure}

Figure~\ref{fig:fs-gain-greedy} shows that \gls{fs} prompting generally improved greedy-decoding performance, with the largest gain of +16.72\% for the \texttt{Llama-3-8B}~\cite{llama3modelcard} group. This improvement was also seen in \texttt{Llama-3.2-3B}~\cite{llama323modelcard} and \texttt{Llama-3.1-8B}~\cite{llama318modelcard} model groups which gained +15.85\% and +8.89\% penalized F1-score on average. In contrast, the \texttt{Llama-3.2-1B}~\cite{llama321modelcard} group not only did not show an improvement, but also had a slight decrease of -0.15\% performance. This result is {\bf surprising}, since \gls{fs} prompting is generally expected to help models better follow the task format and infer the intended label structure from examples~\cite{brown2020language,abdulghaffar2025llmssuitabilitynetworksecurity}. However, similar observations have been reported in prior work by Lin \textit{et al.}~\cite{lin2025large}. One possible explanation in our case is that the 1B models may have limited capacity to effectively use the in-context examples for multi-label \gls{stride} reasoning. This indicates that \gls{fs} prompting can be beneficial~\cite{brown2020language}, but its effectiveness depends on the model's ability to interpret and apply the provided examples.

% \begin{figure}[!t]
%     \centering
%     \includegraphics[width=0.95\columnwidth]{performancegain_greedy.png}
%     \caption{Few-shot Performance Gain under Greedy Decoding}
%     \label{fig:fs-gain-greedy}
% \end{figure}

The effect of chat-template usage was also inconsistent. In some cases, using the model's chat template improved performance, while in other cases it led to the opposite. For instance, the best overall performance in greedy configuration belongs to \texttt{Llama-3.1-8B-Instruct}~\cite{llama318modelcard} with \gls{fs} but without the chat template. Another point worth noting is that prompting without chat template led models to frequently produce explanatory text rather than only outputting the requested \gls{stride} labels, despite the prompt explicitly instructing the models to avoid additional explanations. Invalid outputs were also more common in no-template settings, suggesting that the absence of a chat template can reduce output-format control. Therefore, although chat-template usage did not consistently improve classification performance, it generally supported better instruction adherence and output-format compliance. This indicates that chat templates usage could help reduce invalid responses of the models.

\subsection{Stochastic Sampling}

In the sampling setting, each model configuration was evaluated over 10 independent runs for each threat description. Unlike greedy decoding, stochastic sampling does not always produce the same output across runs. Therefore, this setting allows us to examine not only classification performance, but also the stability of each model's \gls{stride} label selections. As described in Section~III, sampling was performed using \texttt{temperature} = 0.7 and \texttt{top\_p} = 0.9. A \gls{stride} category was selected if it appeared in at least 50\% of the 10 runs. In cases where no category reached this threshold, the category or categories with the highest selection frequency were selected instead, so that each threat received at least one final \gls{stride} classification.

\begin{figure*}[!t]
  \centering
  \includegraphics[width=0.95\textwidth]{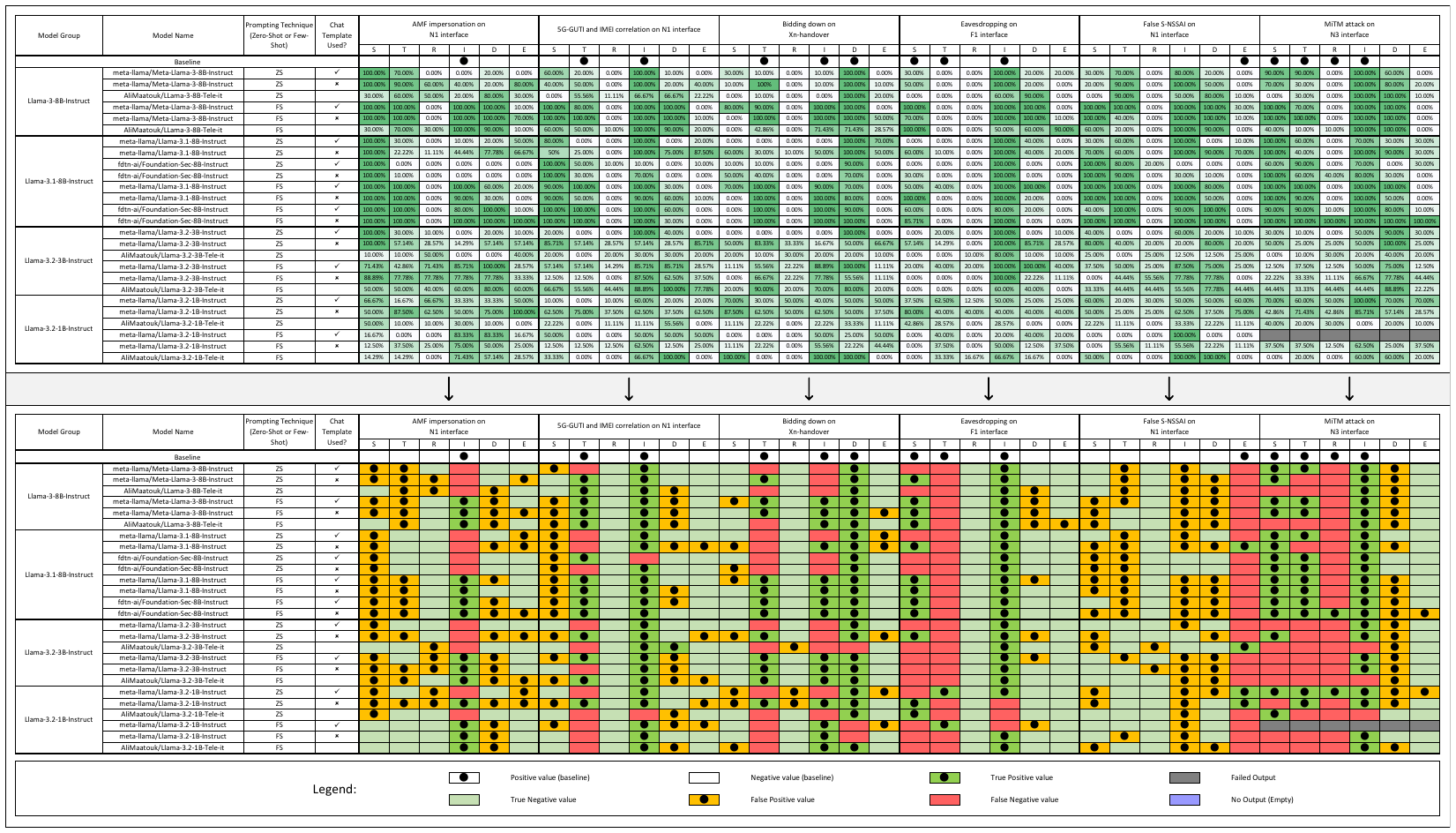}
  \caption{Percentage Distribution and STRIDE Classification of 5G Threats Under Stochastic Sampling.}
  \label{fig:sampling-stride-classification}
\end{figure*}

Figure~\ref{fig:sampling-stride-classification} illustrates how the sampling percentages were converted into final \gls{stride} labels. For each model configuration and threat description, we first calculated the percentage of the 10 runs in which each \gls{stride} category was selected. These percentages were then converted into binary label decisions using the aggregation rule described above: categories selected in at least 50\% of runs were marked as final labels. If no category reached the 50\% threshold, the category or categories with the highest selection frequency were selected instead. Therefore, Figure~\ref{fig:sampling-stride-classification} provides the final label-level classification derived from the sampling percentages.

\begin{table*}[!t]
  \centering
  \caption{Metrics Calculated based on Stochastic Sampling Results \\ \scriptsize{Note: Different cell colors indicate relative metric quality. \textcolor{Green}{\textbf{Green}} represents more desirable values, while \textcolor{Red}{\textbf{Red}} represents less desirable values and white shows neutral results.}}
  \label{tab:sampling-metrics}
  \includegraphics[width=0.95\textwidth]{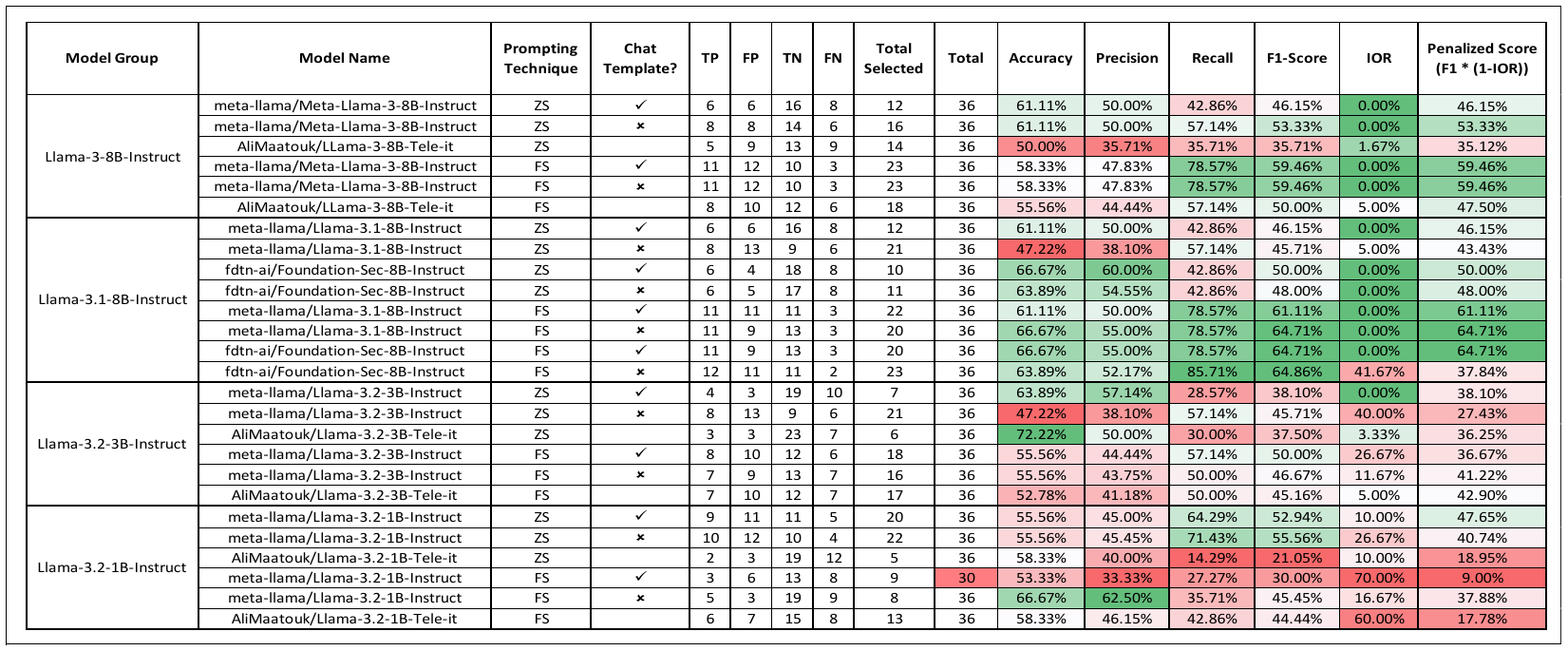}

\end{table*}

\begin{figure}[!t]
    \centering    \includegraphics[width=0.95\columnwidth]{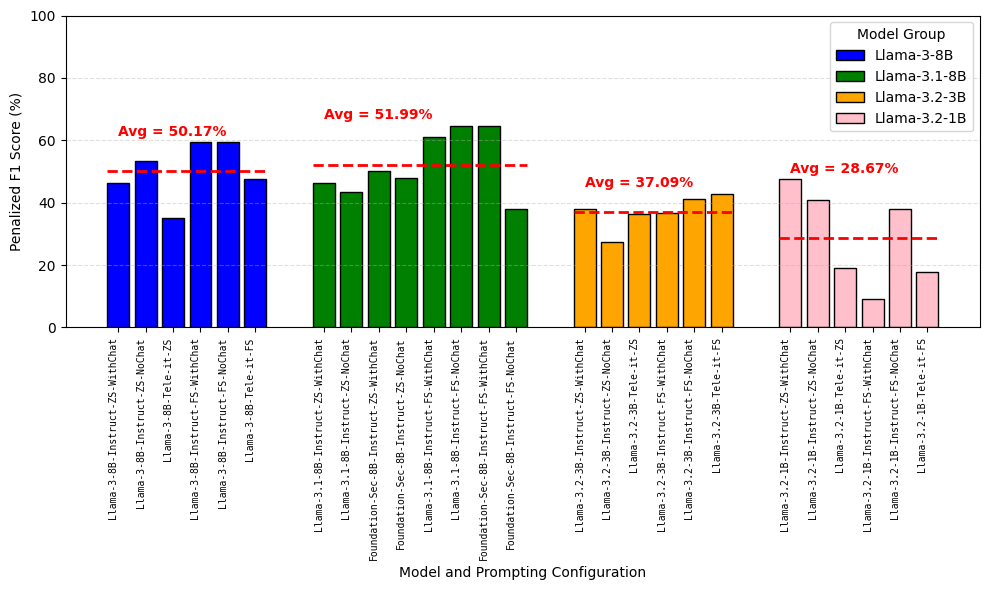}
    \caption{Penalized F1-Score under Stochastic Sampling, Sorted by Model Group}
    \label{fig:f1-sampling}
\end{figure}

Table~\ref{tab:sampling-metrics} reports the metrics calculated from the sampling-based results. Overall, the sampling results show performance patterns that are broadly similar to those observed under greedy decoding. The best sampling results were again generally achieved by 8B-parameter models, especially under \gls{fs} prompting. For example, \texttt{Llama-3.1-8B-Instruct}~\cite{llama318modelcard} with \gls{fs} prompting and no chat template achieved a penalized F1-score of 64.71\%, while \texttt{Foundation-Sec-8B-Instruct}~\cite{weerawardhena2025llama31foundationaisecurityllm8binstructtechnicalreport} with \gls{fs} prompting and chat-template usage also achieved 64.71\%. \texttt{Meta-Llama-3-8B-Instruct}~\cite{llama3modelcard} under \gls{fs} prompting achieved 59.46\% in both chat-template and no-template settings. Figure~\ref{fig:f1-sampling} summarizes the penalized F1-scores per model group under stochastic sampling, showing that larger models achieve higher penalized F1-scores in comparison to the smaller models.

\begin{figure}[!t]
    \centering    \includegraphics[width=0.95\columnwidth]{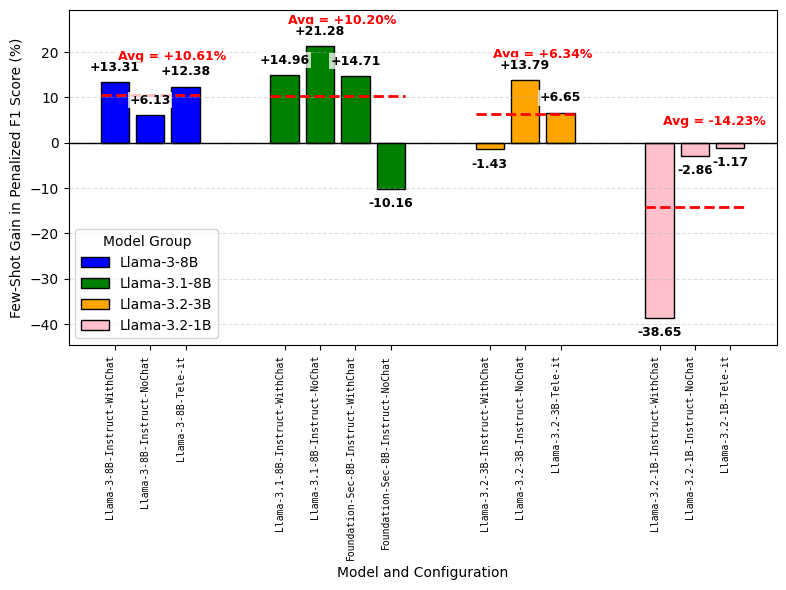}
    \caption{Few-shot Performance Gain under Stochastic Sampling}
    \label{fig:fs-gain-sampling}
\end{figure}

The sampling results also show that \gls{fs} prompting generally improved recall-oriented performance, but the gains were not uniform across all models. Several \gls{fs} settings increased the number of true positives and reduced false negatives compared with their \gls{zs} counterparts. However, in some cases, this improvement was accompanied by additional false positives or a higher \gls{ior}, which reduced the final penalized F1-score. This is particularly important because penalized F1 accounts for both classification quality and output validity. This comparison is shown in Figure \ref{fig:fs-gain-sampling}.

% \begin{figure*}[!t]
%     \centering
%     \includegraphics[width=0.95\textwidth]{F1_SortedByGroup_Sampling.png}
%     \caption{Penalized F1-Score under Sampling, Sorted by Model Group}
%     \label{fig:f1-sampling}
% \end{figure*}

{\bf Compared to greedy decoding}, sampling provides additional information about model uncertainty because it allows the model to generate from a distribution of plausible outputs rather than always selecting the highest-probability next token. Under \texttt{temperature}-based sampling with \texttt{top\_p = 0.9} filtering, labels with lower but still non-negligible probability may be selected across different runs. As a result, some fluctuations are seen in label selection. This variability was especially visible among smaller models. In some 1B and 3B configurations, the model did not consistently select the same categories across runs, leading to lower stability under the 50\% selection threshold. In contrast, the stronger 8B configurations generally produced more stable selections and clearer multi-label classifications.

\subsection{Invalid Output Classification in Greedy and Sampling}
\label{sec:invalid-output}

In addition to classification errors, we also observed several outputs that did not comply with the \gls{stride} classification format. These outputs were all treated as invalid and were not included in traditional metrics (Accuracy, Precision, Recall, F1-score)
computation as we could not reliably convert them to \gls{stride} labels. Instead, we report them separately using the \textbf{\gls{ior}} as can be seen in Tables~\ref{tab:greedy-metrics} and~\ref{tab:sampling-metrics}. Note that penalized F1-score is computed by adjusting the standard F1-score based on \gls{ior}, and therefore the invalid outputs directly impact the final performance score.

\begin{figure}[!t]
    \centering
    \includegraphics[width=0.95\columnwidth]{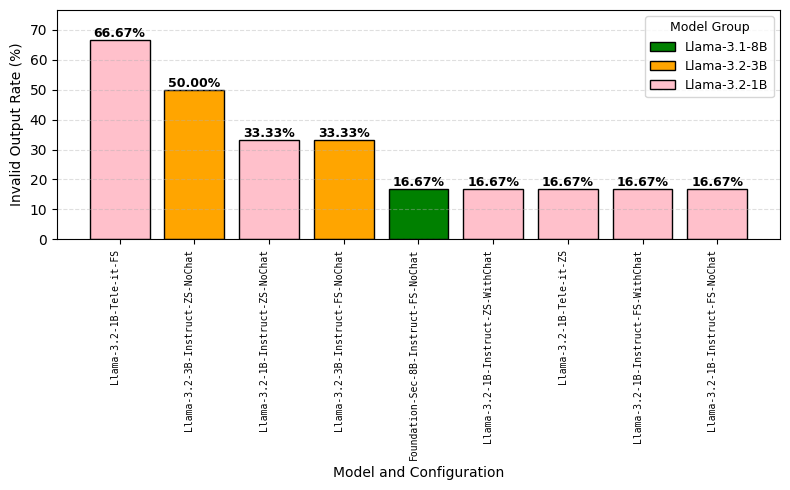}
    \caption{Models' Invalid Output Rate in Greedy Decoding}
    \label{fig:IOR}
\end{figure}

\begin{figure}[!t]
    \centering    
    \includegraphics[width=0.95\columnwidth]{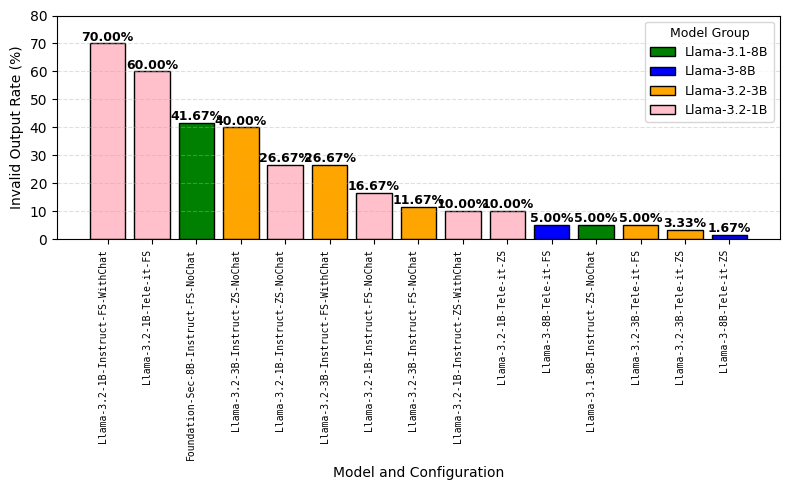}
    \caption{Models' Invalid Output Rate in Stochastic Sampling}
    \label{fig:IOR-sampling}
\end{figure}

Figures~\ref{fig:IOR} and~\ref{fig:IOR-sampling} report the \glspl{ior} across model configurations in greedy and sampling settings, respectively. The figures show that invalid outputs were more common among smaller models and in settings where the model was prompted without its chat template. This suggests that output validity is affected not only by the capability of the model, but also by other factors such as the prompting style.

A qualitative review of the invalid responses revealed several recurring failure patterns. Explanatory text was not considered invalid by itself, provided that the response included a clear \gls{stride} label and that the explanation was consistent with it. However, some outputs provided only a description of the threat or a general explanation of \gls{stride} categories without producing a usable final classification. These outputs were marked invalid because they could not be mapped to the required label set. We also observed cases where the model produced both a \gls{stride} label and an explanation, but the explanation contradicted the predicted label. For instance, a response could explain the threat as spoofing while assigning tampering as the final label. These label-explanation inconsistencies were treated as invalid because the intended classification was unclear. This is an example of what Turpin \textit{et al.}~\cite{turpin2023language} refer to as an unfaithful explanation.

The next group of invalid outputs consists of \textit{label-space violations}. Meaning the model introduced labels or terms outside the six \gls{stride} categories, such as interface names, threat descriptions, or unrelated security concepts. Another recurring pattern was \textit{conflicting or multiple classifications}. Some responses contained more than one classification for the same threat, where the classifications were inconsistent with each other. For example, a model could first provide one set of \gls{stride} labels and then revise or repeat the answer with a different set of labels. These outputs were treated as invalid because they did not provide a single unambiguous prediction.

On a similar note, \textit{missing, partial, or malformed outputs} were also observed during the evaluation process. \textit{Missing} refers to outputs with the word \textbf{none} or completely blank empty outputs. \textit{Partial} and \textit{malformed outputs} represent the outputs that stopped before completing the classification and symbol-only responses that prevented reliable label extraction, respectively. 

The last failure pattern involved unsupported generated content, such as hallucinated non-existent external references and generated code. Such content was not treated as invalid by itself if the response also contained a clear \gls{stride} label that could be extracted. For instance, a response that provided valid \gls{stride} labels followed by an unsupported explanation or reference was still evaluated based on the extracted labels. However, responses consisting only of unsupported explanation, code generation, or unrelated content without a usable \gls{stride} label were marked invalid.

It is also worth noting that none of the models evaluated in our experiments failed gracefully by explicitly expressing uncertainty, abstaining, or indicating that the classification could not be determined. This observation is consistent with prior work on model self-knowledge, which shows that \glspl{llm} still struggle to reliably recognize and communicate the limits of their knowledge, even though instruction tuning, \gls{icl}, and model scale can improve this ability~\cite{yin-etal-2023-large}.

\section{Insights}
\label{sec:insights}
This section synthesizes findings from our greedy-decoding, stochastic-sampling, and invalid-output analyses. Rather than focusing merely on the optimal configuration, we interpret these results to evaluate the practical applicability of current \glspl{llm} for structured \gls{stride}-based threat modeling.

\subsection{RQ1: Domain Adaptation Alone Is Not Sufficient:}
\label{c1}
Across the evaluated configurations, domain-adapted models did not consistently outperform their general-purpose counterparts. Although cybersecurity and telecommunications-adapted models such as \texttt{Foundation-Sec-8B-Instruct}~\cite{weerawardhena2025llama31foundationaisecurityllm8binstructtechnicalreport} and the \texttt{Tele-it}~\cite{maatouk2025telellmsseriesspecializedlarge} variants achieved high penalized F1-scores in some settings, their performance was not always better than the corresponding general-purpose Llama Instruct models. In several cases, general-purpose instruction-tuned models achieved comparable or higher penalized F1-scores, particularly under \gls{fs} prompting.

This suggests that domain-specific pretraining alone is not sufficient for reliable \gls{stride}-based threat classification. One possible explanation is that \gls{stride} classification requires more than recognition of cybersecurity or telecommunications terminology. The model must also reason about the security impact of a threat, map that impact to one or more abstract \gls{stride} categories, and follow a constrained output format. Therefore, task-specific reasoning ability and instruction adherence appear to be as important as domain knowledge. These findings indicate that future domain adaptation efforts for threat modelling should not only expose models to cybersecurity and telecommunications corpora, but also align them with structured threat-modelling tasks and security reasoning.

\newtcolorbox{rqbox1}{
    colback=blue!5,
    colframe=black,
    boxrule=0.5pt,
    arc=1pt,
    left=4pt,
    right=4pt,
    top=4pt,
    bottom=4pt,
    fonttitle=\bfseries,
    before skip=6pt,
    after skip=6pt
}

\begin{rqbox1}
\textbf{Answering RQ1:} The results provide limited evidence that domain adaptation alone improves \gls{stride}-based threat classification. Domain-adapted models achieved high penalized F1-score in some settings, but their gains were inconsistent and they did not frequently outperform their general-purpose counterparts.
\end{rqbox1}

\subsection{RQ2: Model Properties:}
\label{c2}

\subsubsection{Larger Models Are Generally Stronger, but Scaling Is Not a Complete Solution}
The results also show that increasing the model size generally improves \gls{stride} classification performance. In both greedy and sampling settings, the best results in term of penelized F1-score, were achieved by 8B-parameter models. Smaller models, particularly the 1B and 3B variants, showed lower penalized F1-scores and were more likely to produce unstable or invalid outputs.

However, the relationship between model size and performance was not strictly monotonic. Increasing model size improved the overall trend, but it did not eliminate classification errors, invalid outputs, or sensitivity to prompting conditions. This suggests that scaling helps, but does not fully solve the challenges of structured threat modelling. Even larger models may still misclassify secondary \gls{stride} categories, over-select plausible but incorrect labels, or fail to follow the required output format. Therefore, model scale should be viewed as one contributor to performance rather than a guarantee of reliability.

\subsubsection{Stochastic Sampling Reveals Classification Uncertainty}

Unlike greedy decoding, which returns a single highest-probability output, stochastic sampling allows each model to generate multiple independent classifications for the same threat. This setting enabled us to observe how consistently models of different sizes and prompting configurations selected each \gls{stride} category across repeated runs. 

Analysis of sampling outputs showed that smaller models were often less consistent in selecting a label, frequently fluctuating across different categories and even sometimes failing to meet a \(50\%\) confidence threshold for label selection. As noted previously, this behaviour is largely driven by how \texttt{temperature} affects smaller models differently from larger ones. Considering this, stochastic sampling with controlled randomness still represents a valuable technique for uncertainty analysis in threat classification tasks.

\newtcolorbox{rqbox2}{
    colback=blue!5,
    colframe=black,
    boxrule=0.5pt,
    arc=1pt,
    left=4pt,
    right=4pt,
    top=4pt,
    bottom=4pt,
    fonttitle=\bfseries,
    before skip=6pt,
    after skip=6pt
    }

\begin{rqbox2}
\textbf{Answering RQ2:} Larger models generally performed better, with 8B-parameter models achieving the best penalized F1-scores in both settings. However, the improvement was not strictly monotonic. Stochastic sampling further showed that smaller models were less consistent with label selection, making this decoding strategy useful not only for performance evaluation purposes, but also for uncertainty analysis.
\end{rqbox2}

\subsection{RQ3: Prompting Technique:}
\label{c3}

\subsubsection{Few-Shot Prompting Generally Improves Recall-Oriented Classification}
Few-shot prompting generally improved performance, especially by increasing the number of correctly identified \gls{stride} labels. This effect was most visible in the larger model groups, where in-context examples helped the models better infer the expected multi-label structure of the task. In several configurations, \gls{fs} prompting reduced false negatives, leading to improved recall and higher F1-scores.

At the same time, the benefit of this approach was not uniform across all models. Particularly for smaller models, \gls{fs} not only did not improve performance, but also introduced additional false positives or invalid outputs. This suggests that the effectiveness of \gls{icl} depends on the model's capacity to interpret and apply the examples correctly. For \gls{stride} threat classification, \gls{fs} prompting appears useful, but it should not be assumed to improve all models equally.

\subsubsection{Chat Templates Improve Instruction Adherence More Than Classification Accuracy}

The effect of chat-template usage was mixed when measured only by penalized F1-score. In some configurations, using the model's chat template improved performance, while in others the no-template setting achieved higher classification scores. Therefore, chat templates should not be treated as universally beneficial for classification accuracy.

However, the qualitative review showed that chat-template usage helped with instruction adherence and output-format control. Without chat templates, models more frequently produced explanations, extra text, malformed responses, or outputs that did not follow the requested label-only format. This distinction is important: a configuration may achieve a comparable F1-score while still being less reliable from a deployment perspective if it frequently violates formatting requirements.

\newtcolorbox{rqbox3}{
    colback=blue!5,
    colframe=black,
    boxrule=0.5pt,
    arc=1pt,
    left=4pt,
    right=4pt,
    top=4pt,
    bottom=4pt,
    fonttitle=\bfseries,
    before skip=6pt,
    after skip=6pt
    }

\begin{rqbox3}
\textbf{Answering RQ3:} Prompting technique substantially affected both classification quality and output reliability. \gls{fs} prompting generally improved recall by helping models identify more applicable \gls{stride} labels, although the benefit was not uniform across all model sizes. Chat-template usage had a mixed effect on penalized F1-score, but it generally improved instruction adherence and reduced the number of invalid outputs.
\end{rqbox3}

\subsection{Additional Insights: Performance Metrics Do Not Fully Capture Output Reliability}
\label{c4}
A key insight from this study is that classification metrics alone do not fully capture model reliability. Some outputs were not simply incorrect; they were unusable for the \gls{stride} classification task. These included responses with no valid \gls{stride} label, label-space violations, conflicting classifications, malformed outputs, hallucinated references, and generated code without a usable label. In other cases, the model provided a \gls{stride} label but contradicted it in the accompanying explanation.

These failures are particularly important for threat modelling because the output must be both semantically correct and operationally usable. A model that produces a correct label in some cases but frequently generates ambiguous, unsupported, or malformed responses would still require substantial human review. The invalid-output analysis therefore shows that reliability in \gls{llm}-assisted threat modelling depends on both classification performance and output validity.
\newtcolorbox{rqbox4}{
    colback=blue!5,
    colframe=black,
    boxrule=0.5pt,
    arc=1pt,
    left=4pt,
    right=4pt,
    top=4pt,
    bottom=4pt,
    fonttitle=\bfseries,
    before skip=6pt,
    after skip=6pt
    }

\begin{rqbox4}
\textbf{Additional Insights:} The invalid output analysis shows that the evaluated models should not be trusted as autonomous threat-modelling agents, even when a model achieves a reasonable F1-score. Some failures were not simple classification errors, but severe reliability failures, including hallucinated categories, non-existent references, and explanations that contradicted the final \gls{stride} label.
\end{rqbox4}

\vspace{0.5cm}

\section{Prompting suggestions for Threat Modelling}
\label{sec:prompting-guidlines}

Prompt engineering is a critical design consideration in \gls{llm}-assisted threat modelling because the usefulness of the output depends not only on the model's underlying capabilities, but also on how the task is framed, formatted, and constrained. 
Our experimental observations suggest that models are highly prompt-sensitive: small changes in wording, spacing, line breaks, output instructions, or the presence of examples can affect both classification performance and output validity. 
Based on these observations, this section summarizes practical prompting suggestions for applying \glspl{llm} to \gls{stride}-based threat classification.

\subsection{Role Assignment and Task Framing}

The first prompting suggestion is to define the model's role and the task context clearly, which is often referred to as role-play prompting~\cite{kong2024better}.
Assigning the model the role of a domain-aware security expert can help orient the response toward the intended threat modelling setting, especially when the input scenarios involve specialized systems such as \gls{5g} networks. 
However, role assignment alone is not sufficient. 
The prompt should also explicitly define the classification task and restrict the model to the permitted \gls{stride} categories. 
This is important because unconstrained prompts may encourage the model to generate out-of-scope security concepts, informal threat descriptions, or invalid labels that do not belong to the \gls{stride} framework.

\subsection{Separate Classification from Explanation}

In multi-label classification tasks such as \gls{stride}-based threat modelling, it is important to clearly separate the final classification from any explanatory text. 
For automated evaluation, models should be instructed to return only the applicable labels and to avoid explanations unless rationales are being evaluated separately. 
This makes the outputs easier to parse, reduces the risk of formatting errors, and supports more consistent quantitative evaluation. 
It also reduces ambiguity during human review, since the classification can be assessed independently from any supporting reasoning.

\subsection{Suggest Explicit Output Templates}

To support better quantitative evaluation further and prevent hallucinated explanations, the prompt, whether \gls{zs} or \gls{fs}, must provide a clear answer template. 
Based on our \gls{fs} results, providing an explicit template in the input prompt guides the model to produce a more readable and machine-parsable output. This is particularly useful in classification tasks, where outputs must be mapped to a predefined label space for evaluation. 

\subsection{Set Sampling Parameters based on Model Scale}

Sampling parameters should be selected carefully based on both task requirements and the scale of the models. Since models of different sizes may respond differently to \texttt{temperature} changes~\cite{renze-2024-effect,li2025exploring}, under high-temperature settings, smaller models may become unstable or produce invalid labels. 
Furthermore, in this specific task, STRIDE-based classification, the label space is fixed and known in advance. Therefore, the model is not expected to generate creative responses, but rather to correctly understand the threat scenarios, identify 5G interfaces, and map the threat to valid STRIDE labels. 
Also, when stochastic is used, nucleus sampling can also help limit the generation to more likely outputs and thus reduce the number of irrelevant outputs.
Ideally, \texttt{temperature} and \texttt{top\_p} should be calibrated separately for each model size.

\section{Conclusion}
\label{sec:conclusion}
In this study, we evaluate domain-adapted \glspl{llm} along with their general-purpose counterparts on structured threat modelling of \gls{5g} threats. We use 4 model groups, 8 models, 2 prompting strategies (\gls{zs} and \gls{fs}), with and without a chat template, resulting in 52 unique model configurations. 

This extensive evaluation provides valuable insights that domain adaptation alone does not guarantee better performance. Moreover, structured threat modelling properties (e.g. specific STRIDE categories), the training/tuning and prompting could be tailored to this specific problem to enhance the outcome. 

Also, performance metrics and reliability need to be both taken into consideration if this approach is to be explored further.

\section{Ethical Considerations}
This research evaluates the domain-adapted \glspl{llm} trained on telecommunications and cybersecurity
data and their general counterparts on \gls{stride} threat modelling of \gls{5g} threats. All language models used in this work are publicly available and open-source. %This work does not involve interactions with human subjects, data derived from human subjects, nor does it identify any high-impact vulnerability in networks and systems.

\section*{Acknowledgment}
This work was supported in part by the Natural Sciences and Engineering Research Council of Canada (NSERC). This research was enabled in part by support provided by Research Computing Services (https://carleton.ca/rcs) at Carleton University.

\bibliographystyle{IEEEtran}
\bibliography{references} % e.g., references.bib

\newpage

\section{Appendix: Failed LLM Outputs}	
\label{sec:appendix}

\begin{figure}[!h]
    \centering
    \includegraphics[width=\linewidth]{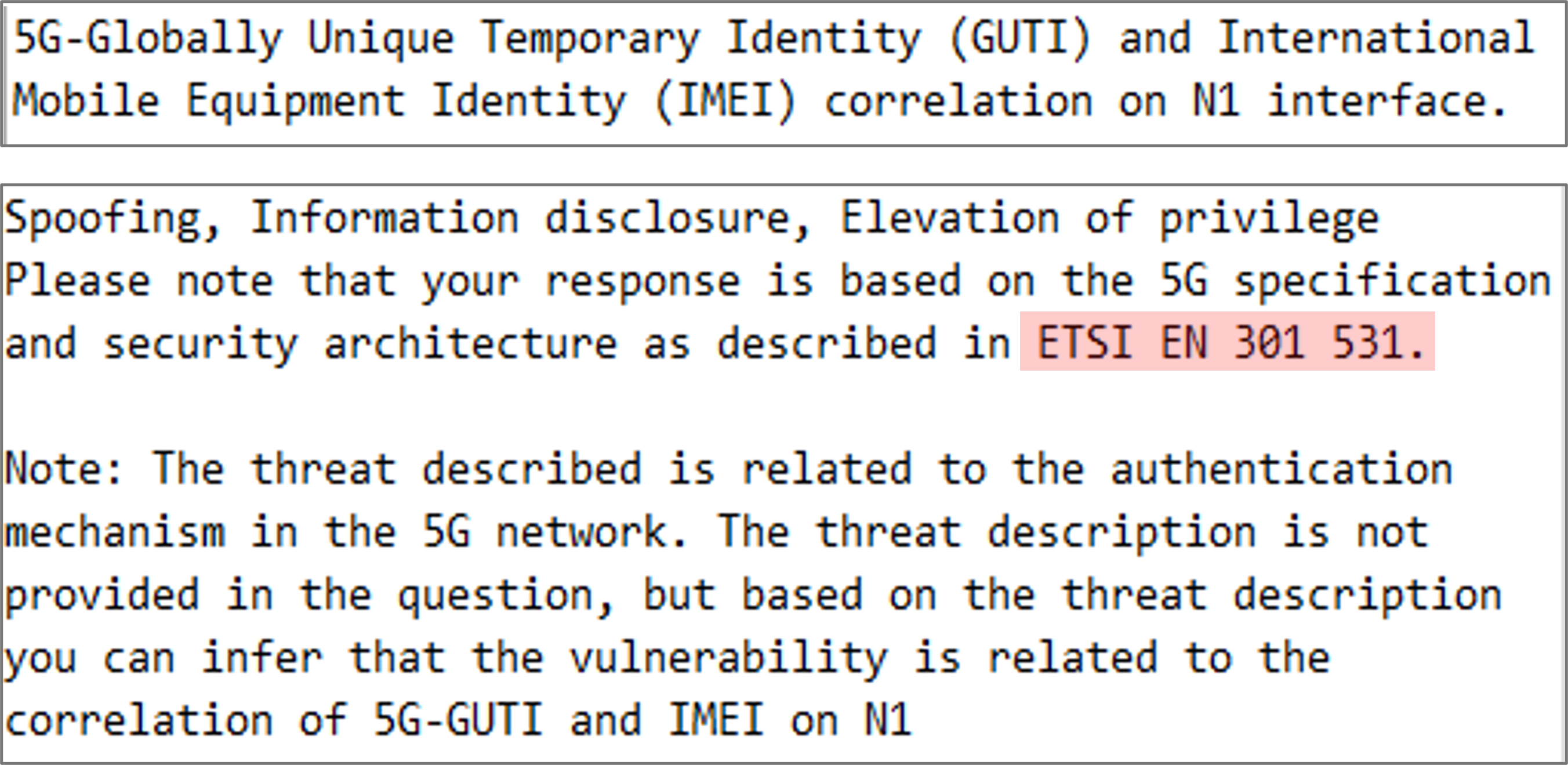}
    \caption{Example of a model providing a reference we could not find.}
    \label{fig:hallucinated-reference}
\end{figure}

Figure~\ref{fig:hallucinated-reference} shows a model providing a reference that we could not find (likely a non-existent standard) to justify its label classification. This type of output is particularly concerning. In practical settings, users may be more likely to trust such outputs when they appear to be supported by an official standard, even when the referenced standard may not exist.

\begin{figure}[!h]
    \centering
    \includegraphics[width=\linewidth]{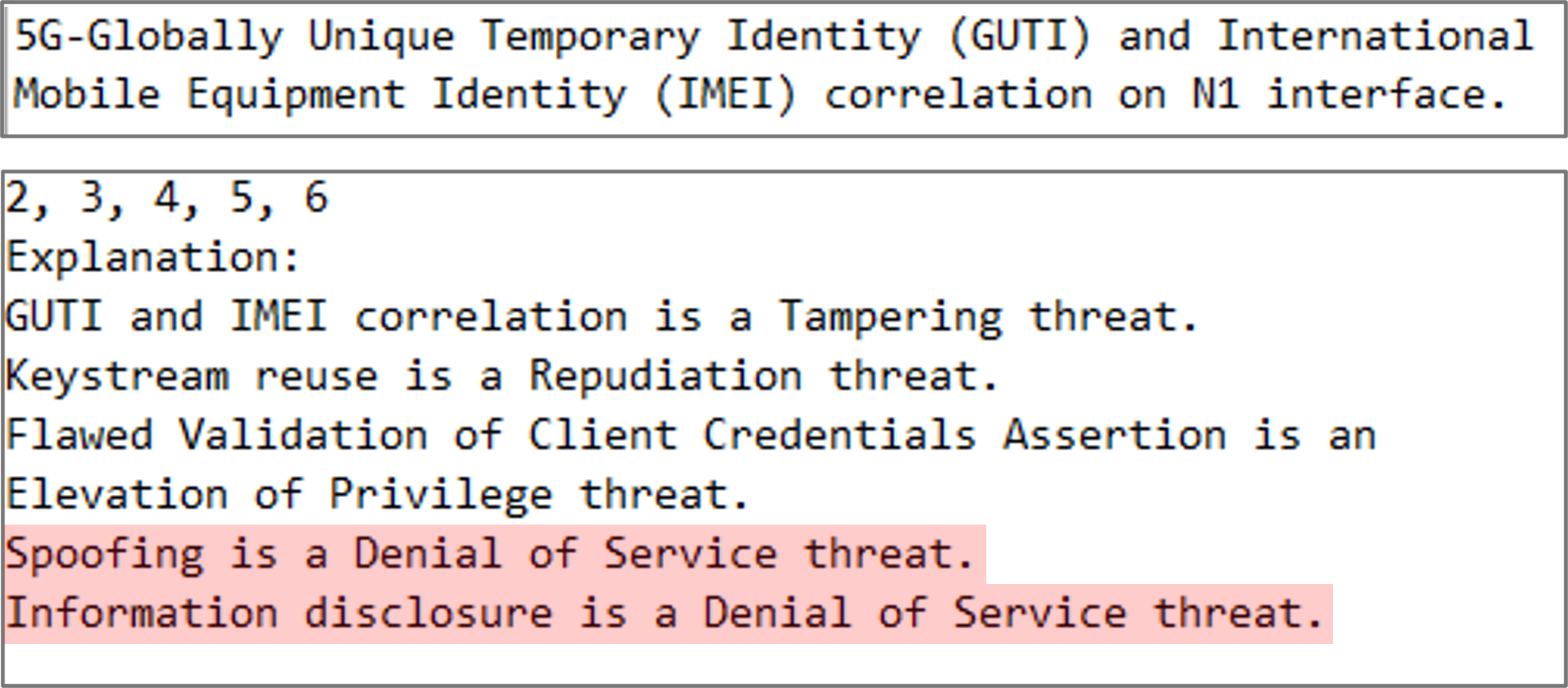}
    \caption{Example of an invalid output where the model provides malformed labels and contradictory STRIDE explanations.}
    \label{fig:malformed-output}
\end{figure}

Figure~\ref{fig:malformed-output} displays another error type, which includes a semantically incorrect explanation of STRIDE labels. This raises an important question: do \glspl{llm} genuinely understand the semantic distinctions between STRIDE categories?

% A single valid run is expected to output one single \gls{stride} classification for each threat. Figure~\ref{fig:nine} shows a model producing nine different classifications for a single threat in a single run. Therefore, none of the given answers could be confidently marked as predicted labels, and this was counted as an invalid output.

% \begin{figure}
%     \centering
%     \includegraphics[width=\linewidth]{NineClassifications.png}
%     \caption{Example of a model outputting nine different classifications for one single threat.}
%     \label{fig:nine}
% \end{figure}

\begin{figure}[!h]
    \centering
    \includegraphics[width=\linewidth]{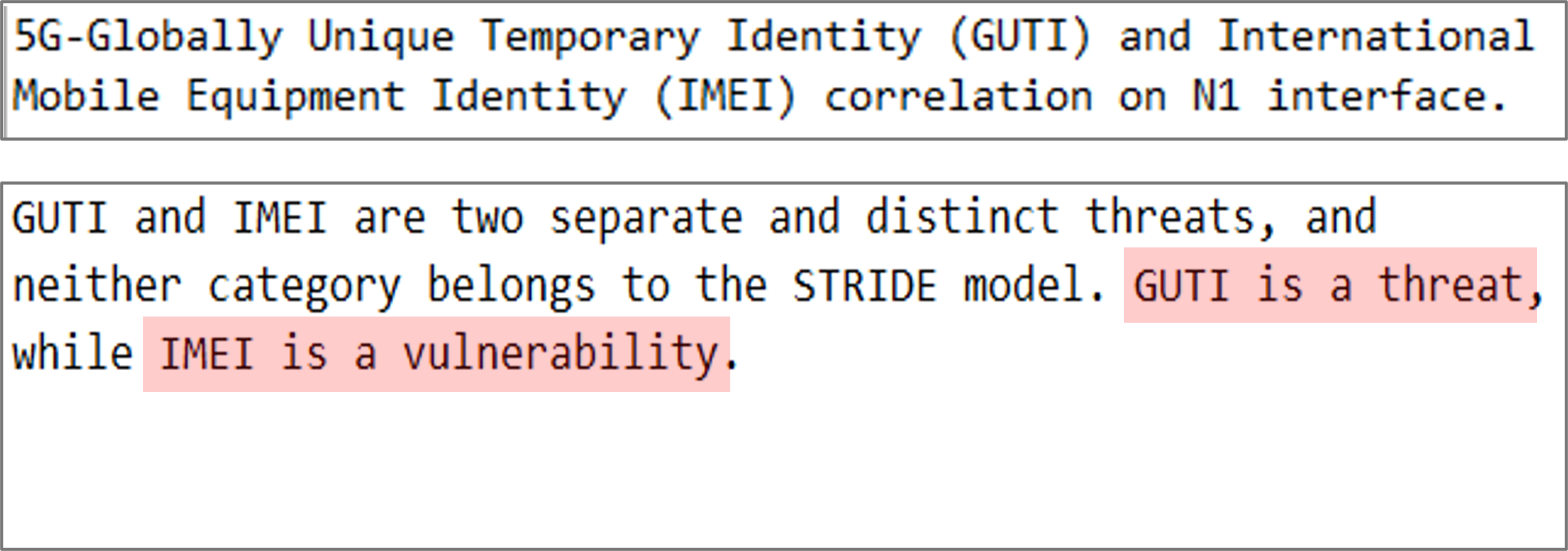}
    \caption{Example of an invalid output where the model incorrectly identifies the UE identifiers (GUTI and IMEI) as distinct threats.}
    \label{GUTIIMEI}
\end{figure}

Figure~\ref{GUTIIMEI} provides another example of invalid output where the model incorrectly identifies the \gls{ue} identifiers (GUTI and IMEI) as distinct threats, rather than providing a \gls{stride} classification of this threat. The model goes on further to classify GUTI as a threat and IMEI as a vulnerability, clearly showcasing its inability to adhere to the user prompt.

\end{document}